\documentstyle[aps,prl,epsf,graphicx]{revtex}
\begin{document}

\twocolumn[\hsize\textwidth\columnwidth\hsize\csname
@twocolumnfalse\endcsname

\title{Gravitational collapse on the brane: a no-go theorem}

\author{Marco Bruni, Cristiano Germani and
Roy Maartens}

\address{
Relativity and Cosmology Group, School of Computer Science and
Mathematics, Portsmouth University, Portsmouth~PO1~2EG, Britain}

\maketitle

\begin{abstract}
We investigate how braneworld gravity affects gravitational
collapse and black hole formation by studying
Oppenheimer-Snyder-like collapse on a Randall-Sundrum type brane.
Without making any assumptions about the bulk, we prove a no-go
theorem: the exterior spacetime on the brane cannot be static,
which is in stark contrast with general relativity. We also
consider the role of Kaluza-Klein energy density in collapse,
using a toy model.

\end{abstract}


\vskip1pc]

The study of gravitational collapse in general relativity (GR) is
fundamental to understanding the behaviour of the theory at high
energies. The Oppenheimer-Snyder (OS) model still provides a
paradigmatic example that serves as a good qualitative guide to
the general collapse problem in GR. It can be solved analytically,
as it simply assumes a collapsing homogeneous dust cloud of finite
mass and radius, described by a Robertson-Walker metric and
surrounded by a vacuum exterior. In GR, this exterior is
necessarily static and given by the Schwarzschild
solution~\cite{s}. In other theories of gravity that differ from
GR at high energies, it is natural to look for similar examples.
Braneworld scenarios of Randall-Sundrum type~\cite{rs,sms} lead to
modified Einstein equations as the effective 4D field equations on
the brane. In this Letter we analyze an OS-like collapse  in this
setting, in order to shed light on some fundamental differences
between collapse in GR and on the brane.

In string theory and M-theory, which may provide a route towards
quantum gravity, gravity is a truly higher-dimensional
interaction, which becomes effectively 4D at low enough energies.
Simple braneworld models inspired by these theories describe the
observable universe as a 3-brane boundary of a 4D space (the
bulk), with matter fields confined on the brane. Gravity
propagates in all 4 spatial dimensions, but must be localized near
the brane at low energies, in order to reproduce the successful
predictions of GR. This is most obviously achieved via a small
compact extra dimension, as in Kaluza-Klein (KK) theories, but the
Randall-Sundrum model~\cite{rs} localizes gravity by the curvature
of the bulk, even with a noncompact extra dimension. The bulk
metric, which is anti-de\,Sitter (AdS$_5$), satisfies the
5-dimensional Einstein equations with negative cosmological
constant. Their model, with vacuum Minkowski brane, has been
generalized to allow for arbitrary energy-momentum tensor on the
brane, and the effective field equations on the brane are modified
Einstein equations~\cite{sms}.

Perturbative analysis of the gravitational field due to a compact
source on the brane has been performed~\cite{rs,ssm,pert}. In
particular, in the weak-field limit there is a 5D correction to
the Newtonian potential on the brane which to leading order gives
\begin{equation}\label{pert}
\Phi=(GM/ r)\left[1+2\ell^2/3r^2\right]\,,
\end{equation}
where $\ell$ is the curvature scale of AdS$_5$. Brane solutions of
static black hole and stellar exteriors with 5-dimensional
corrections to the Schwarzschild metric have been
found~\cite{dmpr,gm}, but the bulk metric for these solutions has
not been found. The Schwarzschild black string bulk metric has a
Schwarzschild black hole on the brane, but cannot describe the end
state of gravitational collapse~\cite{chr}. Numerical integration
into the bulk, starting from static black hole solutions on the
brane, is plagued with difficulties~\cite{num}. In summary, very
little is known about astrophysical black holes and stars on the
brane, even in the static case. Even less is known about
astrophysical gravitational collapse on the brane to a black hole.

Braneworld gravitational collapse is complicated by a number of
factors. The confinement of matter to the brane, while the
gravitational field can access the extra dimension, is at the root
of the difficulties relative to Einstein's theory, and this is
compounded by the gravitational interaction between the brane and
the bulk. Matching conditions on the brane are more complicated to
implement~\cite{gm}, and one also has to impose regularity and
asymptotic conditions on the bulk, and it is not obvious what
these should be.

In GR, the OS model of collapsing dust has a Robertson-Walker
interior matched to a Schwarzschild exterior. We show that even
this simplest case is much more complicated on the brane. However,
it does have a striking new property, which may be part of the
generic collapse problem on the brane. The exterior is not
Schwarzschild, and nor could we expect it to be, as discussed
above, but the exterior is not even {\em static}, as shown by our
no-go theorem. The reason for this lies in the nature of the
braneworld modifications to GR.

The field equations on a generalized Randall-Sundrum brane
are~\cite{sms}
\begin{equation}\label{fe}
G_{\mu\nu}=-\Lambda g_{\mu\nu}+ 8\pi GT_{\mu\nu}+48\pi G
S_{\mu\nu}/\lambda-{\cal E}_{\mu\nu}\,,
\end{equation}
where $\lambda$ is the brane tension ($\lambda>
10^8~\mbox{GeV}^4$) and the brane cosmological constant $\Lambda$
has contributions from the tension and the bulk cosmological
constant. The tensor $S_{\mu\nu}$ is quadratic in the
energy-momentum tensor $T_{\mu\nu}$, and ${\cal E}_{\mu\nu}$ is
the projection of the bulk Weyl tensor. The energy-momentum tensor
satisfies the usual conservation equations, and the Bianchi
identities on the brane then imply a ``conservation" equation for
the tracefree ${\cal E}_{\mu\nu}$:
\begin{equation}\label{con}
\nabla^\nu T_{\mu\nu}=0\,,~\nabla^\nu{\cal E}_{\mu\nu}=48\pi G
\nabla^\nu S_{\mu\nu}/\lambda\,,~{\cal E}^\mu{}_\mu=0.
\end{equation}
The general 1+3 form of these equations is given in~\cite{m}.
Five-dimensional corrections to the field equations of GR are of
two types~\cite{sms,m}.

\noindent {\bf (1)}~\underline{High-energy corrections}, given by
$S_{\mu\nu}$, arise from the extrinsic curvature of the brane, and
increase the effective density and pressure of collapsing matter.
In particular this means that the effective pressure does not in
general vanish at the surface, changing the nature of the matching
conditions on the brane~\cite{gm}. Gravitational collapse
inevitably produces energies high enough to make these corrections
significant.

\noindent {\bf (2)}~\underline{KK corrections}, given by ${\cal
E}_{\mu\nu}$, arise from 5D graviton stresses, and are constrained
by Eq.~(\ref{con}). In the linearized regime they are known as
massive KK modes of the graviton~\cite{rs}; in general, they are a
signature of nonlinear KK modes in the bulk. For brane-bound
observers, these stresses are nonlocal: local density
inhomogeneities on the brane generate Weyl curvature in the bulk
that ``backreacts" nonlocally on the brane~\cite{m}. Even in the
absence of matter, ${\cal E}_{\mu\nu}$ may be nonzero (provided
that $\nabla^\nu{\cal E}_{\mu\nu}=0$), since there may be 5D Weyl
curvature in the bulk, e.g. sourced by a bulk black hole, as in
cosmological braneworld models~\cite{msm,bcg}. The KK stresses
further complicate the matching problem on the brane, since they
in general contribute to the effective radial pressure at the
surface.

Equations~(\ref{fe}) and (\ref{con}) are the complete set of
equations on the brane. They are not closed, since ${\cal
E}_{\mu\nu}$ contains 5D degrees of freedom that cannot be
determined on the brane. A further set of 5D equations~\cite{sms}
makes up the full closed system. However, using only the 4D
projected equations, we prove a no-go theorem valid for the full
5D problem: {\em given the standard matching conditions on the
brane, the exterior of a collapsing dust cloud cannot be static}.
We are not able to determine the non-static exterior metric, but
we expect on general physical grounds that the non-static
behaviour will be transient, so that the exterior tends to a
static form.

The collapsing region in general contains dust and also energy
density on the brane from KK stresses in the bulk (this is called
``dark radiation" in cosmology~\cite{msm,bcg}). We show that in
the extreme case where there is no matter but only collapsing
homogeneous KK energy density, there is a unique exterior which is
static for physically reasonable values of the parameters. Since
there is no matter on the brane to generate KK stresses, the KK
energy density on the brane must arise from bulk Weyl curvature.
In this case, the bulk could be pathological. The collapsing KK
energy density can either bounce or form a black hole with a 5D
gravitational potential, and the exterior is of the Weyl-charged
de\,Sitter type (given in~\cite{dmpr} for $\Lambda=0$), but with
no mass.

The collapse region has a Robertson-Walker metric
\begin{equation}\label{1}
ds^2=-d\tau^2+a(\tau)^2(1+ {\textstyle {1\over4}}
kr^2)^{-2}\left[dr^2+r^2d\Omega^2\right]\,.
\end{equation}
The modified Friedmann equation from Eq.~(\ref{fe}) is~\cite{bdel}
\begin{eqnarray} \label{evol}
{\dot{a}^2}/{a^2}= {\textstyle{8\over3}}\pi G\rho\left(1 +
\rho/2\lambda\right) + {C}/{\lambda a^4} -{k}/{a^2}+
{\textstyle{1\over3}} {\Lambda}\,,
\end{eqnarray}
where the KK constant $C$ is fixed by the bulk Weyl curvature (for
a cosmological Friedmann brane, $C$ is proportional to the mass of
a black hole in the {\em bulk}~\cite{msm,bcg}). The $\rho^2$ term,
which is significant for $\rho\gtrsim \lambda$, is the high-energy
correction term, following from $S_{\mu\nu}$. Standard Friedmann
evolution is regained in the limit $\lambda^{-1} \to 0$.
Eq.~(\ref{con}) implies $\rho=\rho_0 (a_0/a)^3$, where $a_0$ is
the epoch when the cloud started to collapse. The proper radius
from the centre of the cloud is $R(\tau)=r a(\tau)/(1+ {\textstyle
{1\over4}} kr^2)$. The collapsing boundary surface $\Sigma$ is
given in the interior comoving coordinates as a free-fall surface,
i.e.\ $r=r_0=$~const, so that $R_\Sigma(\tau)= r_0 a(\tau)/(1+
{\textstyle {1\over4}} kr_0^2)$.

We can rewrite the modified Friedmann equation on the interior
side of $\Sigma$ as
\begin{equation}\label{geo1}
\dot{R}^2= {2GM}/{R}+ 3{GM^2}/{4\pi\lambda R^4}+ {Q}/{\lambda
R^2}+ E+ {\Lambda}/3 R^2\,,
\end{equation}
where the ``physical mass" $M$ (total energy per proper star
volume) and the total ``tidal charge" $Q$ are
\begin{equation}\label{mq}
M={\textstyle{4\over3}}\pi a_0^3 r_0^3\rho_0(1+ {\textstyle
{1\over4}}kr_0^2)^{-3}, Q=C{r_0^4}(1+ {\textstyle
{1\over4}}kr_0^2)^{-4},
\end{equation}
and the ``energy" per unit mass is given by
\begin{equation}\label{en}
E=-{kr_0^2}(1+ {\textstyle {1\over4}}kr_0^2)^{-2}>-1 \,.
\end{equation}

Now we assume that the exterior is static, and satisfies the
standard 4D junction conditions. Then we check whether this
exterior is physical by imposing the modified Einstein
equations~(\ref{fe}) for vacuum, i.e.\ for $T_{\mu\nu} =0 =
S_{\mu\nu}$. The standard 4D Israel matching conditions, which we
assume hold on the brane, require that the metric and the
extrinsic curvature of $\Sigma$ be continuous. The extrinsic
curvature is continuous if the metric is continuous and if $\dot
R$ is continuous~\cite{s}. We therefore need to match the metrics
and $\dot R$ across $\Sigma$.

The most general static spherical metric that could match the
interior metric on $\Sigma$ is
\begin{eqnarray}
ds^2&=& -F(R)^2A(R)dt^2 +{dR^2}/A(R) +R^2d\Omega^2\,, \nonumber\\
A(R) &=&1-2Gm(R)/R \,.\label{s1}
\end{eqnarray}
We need two conditions to determine the functions $F(R)$ and
$m(R)$. Now $\Sigma$ is a freely falling surface in both metrics,
and the radial geodesic equation for the exterior metric gives
$\dot{R}^2=-A(R)+{\tilde{E}/ F(R)^2}\,,$ where $\tilde{E}$ is a
constant and the dot denotes a proper time derivative, as above.
Comparing this with Eq.~(\ref{geo1}) gives one condition. The
second condition is easier to derive if we change to null
coordinates. The exterior static metric, with
$dv=dt+dR/[F(1-2Gm/R)]$, becomes
\begin{eqnarray}
ds^2&=& -F^2Adv^2 +2FdvdR+R^2d\Omega^2\,.\label{s1'}
\end{eqnarray}
The interior Robertson-Walker metric takes the form~\cite{tesi}
\begin{eqnarray}
ds^2&=& -\tau_{,v}^2\left[ 1- (k+\dot{a}^2)R^2/a^2 \right] dv^2/
(1-kR^2/a^2) \nonumber\\&&~{} +2 \tau_{,v} dvdR/\sqrt{1-kR^2/a^2}
+R^2d\Omega^2\,,\label{s1''}
\end{eqnarray}
where $d\tau=\tau_{,v}dv+(1+{1\over4}kr^2) dR/[r\dot
a-1+{1\over4}kr^2]$. Comparing Eqs.~(\ref{s1'}) and (\ref{s1''})
on $\Sigma$ gives the second condition. The two conditions
together imply that $F$ is a constant, which we can take as
$F(R)=1$ without loss of generality (choosing $\tilde E=E+1$), and
that
\begin{equation}\label{s3}
m(R)= M+{3M^2}/{8\pi\lambda R^3}+ {Q}/{2G\lambda R}+ {\Lambda
R^3}/{6G}\,.
\end{equation}
In the limit $\lambda^{-1}\to 0$, we recover the 4D GR
Schwarzschild-de\,Sitter solution. However, we note that the above
form of $m(R)$ violates the weak-field perturbative limit in
Eq.~(\ref{pert}), and this is a symptom of the problem with a
static exterior. Equations~(\ref{s1}) and (\ref{s3}) imply that
the brane Ricci scalar is
\begin{eqnarray}\label{m1}
R^\mu{}_\mu=4\Lambda+{9GM^2}/ {2\pi\lambda R^6}\,.
\end{eqnarray}
However, Eq.~(\ref{fe}) for a vacuum exterior implies
\begin{equation}\label{vac}
R_{\mu\nu}=\Lambda g_{\mu\nu}- {\cal E}_{\mu\nu}\,,~R^\mu{}_\mu= 4
\Lambda\,.
\end{equation}
Comparing with Eq.~(\ref{m1}), we see that a static exterior is
only possible if $M/\lambda=0\,.$ This is obviously satisfied in
the GR limit, but {\em in the braneworld, collapsing homogeneous
and isotropic dust leads to a non-static exterior.} We emphasize
that this no-go result does not require any assumptions on the
nature of the bulk spacetime.

The one case that escapes the no-go theorem is $M=0$.  In GR,
$M=0$ would lead to vacuum throughout the spacetime, but in the
braneworld, there is the tidal KK stress on the brane, i.e.\ the
$Q$-term in Eq.~(\ref{geo1}).  The possibility of black holes
forming from KK energy density was suggested in~\cite{dmpr}. The
dynamics of a Friedmann universe (i.e.\ without exterior),
containing no matter but only KK energy density (``dark
radiation") has been considered in~\cite{bcg}. In that case, there
is a black hole in the Schwarzschild-AdS$_5$ bulk, which sources
the KK energy density. Growth in the KK energy density corresponds
to the black hole and brane moving closer together; a singularity
on the brane can arise if the black hole meets the brane. Here we
investigate the collapse of a bound region of homogeneous KK
energy density within an inhomogeneous exterior. It is not clear
whether the bulk black hole model may be modified to describe this
case, and we do not know what the bulk metric is. However, we know
that there must be 5D Weyl curvature in the bulk, and that the
bulk could be pathological, with a more severe singularity than
Schwarzschild-AdS$_5$. Even though such a bulk would be unphysical
(as in the case of the Schwarzschild black string), it is
interesting to explore the properties of a brane with collapsing
KK energy density, since this idealized toy model may lead to
important physical insights into more realistic collapse with
matter and KK energy density.

The exterior is static and unique, and given by the Weyl-charged
de\,Sitter metric
\begin{equation}\label{wd}
ds^2=-Adt^2+{dR^2/ A}+R^2d\Omega^2\,,~M=0\,,
\end{equation}
if $A>0$. For $Q=0$ it is de\,Sitter, with horizon
$H^{-1}=\sqrt{3/\Lambda}$. For $\Lambda=0$ it is the special case
$M=0$ of the solutions given in~\cite{dmpr}, and the length scale
$H_Q^{-1}=\sqrt{|Q|/\lambda}$ is an horizon when $Q>0$; for $Q<0$,
there is no horizon. As we show below, the interplay between these
scales determines the characteristics of collapse.

For $\Lambda=0$, the exterior gravitational potential is
\begin{equation}
\Phi=Q/2\lambda R^2\,,
\end{equation}
which has the form of a purely 5D potential when $Q>0$. When
$Q<0$, the gravitational force is repulsive. We thus take $Q>0$ as
the physically more interesting case, corresponding to {\em
positive} KK energy density in the interior. However we note the
remarkable feature that $Q>0$ also implies {\em negative} KK
energy density in the exterior:
\begin{equation}
-{\cal E}_{\mu\nu}u^\mu u^\nu=\left\{ \begin{array}{ll}
+3Q/(\lambda R_\Sigma^4)\,, & R<R_\Sigma\,, \\ 
-Q/(\lambda R^4)\,, & R> R_\Sigma\,. \end{array}\right.
\end{equation}
Negativity of the exterior KK energy density in the general case
with matter has been previously noted~\cite{ssm,dmpr}.

The boundary surface between the KK ``cloud" and the exterior has
equation of motion $\dot{R}^2=E-V(R)$, where $V=A-1$. For
$\Lambda=0$, the cases are:

\noindent \underline{$Q>0$, $\Lambda=0\,$}:\, The cloud collapses
for all $E$, with horizon at $R_{\rm
h}=H_Q^{-1}=\sqrt{Q/\lambda}$. For $E<0$, given that $E>-1$, the
collapse can at most  start from rest at $R_{\rm
max}=\sqrt{Q/(\lambda|E|)}>R_{\rm h}$.

\noindent \underline{$Q<0$, $\Lambda=0\,$}:\, It follows that
$E>0$, there is no horizon, and the cloud bounces at $R_{\rm
min}=\sqrt{|Q|/(\lambda E)}$.

For $\Lambda> 0$, the potential is given by $V/ V_{\rm c}=-(R/
R_{\rm c})^2\left[1+\epsilon(R_{\rm c}/ R)^4\right]\,,$ where
$V_{\rm c}={H/ H_Q}\,,$ $R_{\rm c}=1/\sqrt{HH_Q}\,,$ and
$\epsilon={\rm sgn}\,Q$ (see~Fig.~1). The horizons are given by
\begin{equation}
\left(R_{\rm h}^\pm\right)^2= R_{\rm c}^2 \left[1\pm
\sqrt{1-4\epsilon V_{\rm c}^2}\right]/2V_{\rm c}\,.\label{hor}
\end{equation}
If $\epsilon>0$ there may be two horizons; then $R_{\rm h}^-$ is
the black hole horizon and $R_{\rm h}^+$  is a modified de\,Sitter
horizon. When they coincide the exterior is no longer static, but
there is a black hole horizon. If $\epsilon<0$ there is always one
de\,Sitter-like horizon, $R_{\rm h}^+$.

\begin{figure}[t]
\begin{center}
\includegraphics[width=8cm,height=4cm,angle=0]{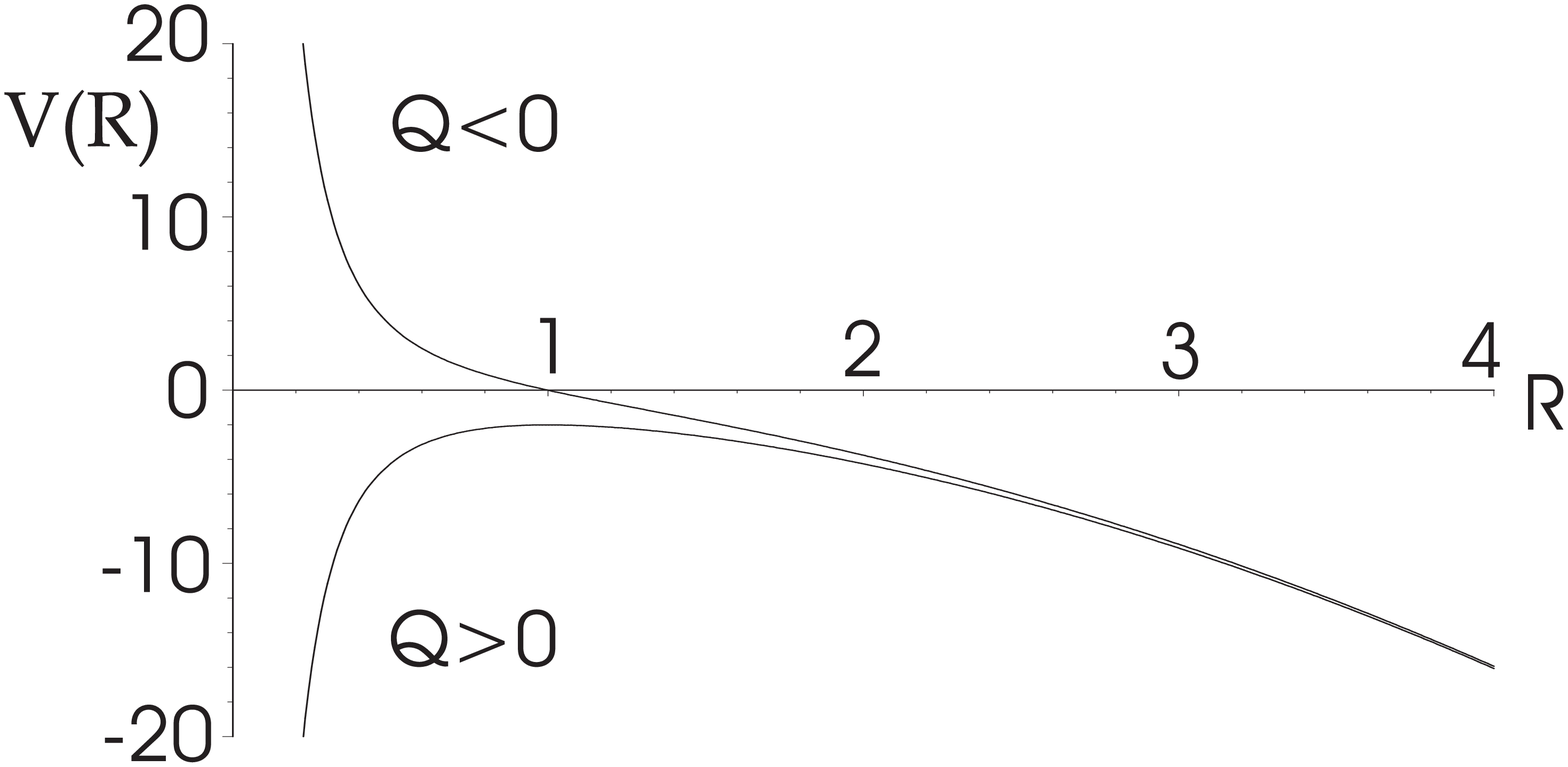}
\end{center}
\label{f3} \caption{The potential $V(R)$ for $\Lambda> 0$, with
$R$ given in units of $R_{\rm c}$ and $V$ given in units of
$V_{\rm c}$.}
\end{figure}

\noindent \underline{$Q>0$, $\Lambda>0\,$}:\, The potential has a
maximum $-2V_{\rm c}$ at $R_{\rm c}$. If $E>-2V_{\rm c}$ the cloud
collapses to a singularity. For $V_{\rm c}>{1\over2}$, i.e.\
$Q>3\lambda/4\Lambda$, there is no horizon, and a naked
singularity forms. For for $V_{\rm c}={1\over2}$ there is one
black hole horizon $R_{h}^{-}=R_{h}^{+}=H^{-1}/\sqrt{2}$. If
$E\leq -2V_{\rm c}$, then Eq.~(\ref{en}) implies $V_{\rm
c}<{1\over2}$, so there are always two horizons in this case.
Either the cloud collapses from infinity down to $R_{\rm min}$ and
bounces, with $R_{\rm min}<R_{\rm h}^+$ always, or it can at most
start from rest at $R_{\rm max} (>R_{\rm h}^-)$, and collapses to
a black hole, where ($\epsilon=1$)
\begin{eqnarray}
R^2_{\stackrel{\scriptstyle \rm min}{\rm max}} =R_{\rm c}^2
\left[-E \pm \sqrt{E^2- 4\epsilon V_{\rm c}^2}\right]/2V_{\rm c}
\,.\label{min}
\end{eqnarray}

\noindent \underline{$Q<0$, $\Lambda>0\,$}:\, The potential is
monotonically decreasing, and there is always an horizon, $R_{\rm
h}^+$. For all $E$, the cloud collapses to $R_{\rm min} (<R_{\rm
h}^+)$, and then bounces, where $R_{\rm min}$ is given by
Eq.~(\ref{min}) with $\epsilon=-1$.

\underline{In summary}, we have explored the consequences for
gravitational collapse of braneworld gravity effects, using the
simplest possible model, i.e.\ an OS-like collapse on a
generalized Randall-Sundrum type brane. Even in this simplest
case, extra-dimensional gravity introduces new features. Using
only the projected 4D equations, we have shown, independent of the
nature of the bulk, that the exterior vacuum on the brane is
necessarily {\em non-static}. This contrasts strongly with GR,
where the exterior is a static Schwarzschild spacetime. Although
we have not found the exterior metric, we know that its non-static
nature arises from (a)~5D bulk graviton stresses, which transmit
effects nonlocally from the interior to the exterior, and (b)~the
non-vanishing of the effective pressure at the boundary, which
means that dynamical information on the interior side can be
conveyed outside. Our results suggest that gravitational collapse
on the brane may leave a signature in the exterior, dependent upon
the dynamics of collapse, so that astrophysical black holes on the
brane may in principle have KK hair.

We expect that the non-static exterior will be transient and {\em
non-radiative}, as follows from a perturbative study of non-static
compact objects, showing that the Weyl term ${\cal E}_{\mu\nu}$ in
the far-field region falls off much more rapidly than a radiative
term~\cite{ssm}. It is reasonable to assume that the exterior
metric will   be static at late times and tend to Schwarzschild,
at least  at large distances.

We have analyzed the idealized collapse of homogeneous KK energy
density whose exterior is static and has purely 5D gravitational
potential. The collapse can either come to a halt and bounce, or
form a black hole or a naked singularity, depending on the
parameter values. This may be seen as a limiting idealization of a
more general spherically symmetric but inhomogeneous case. The
case that includes matter may be relevant to the formation of
primordial black holes in which nonlinear KK energy density could
play an important role.

CG is supported by a PPARC studentship. MB and RM thank the
Mathematical Cosmology Programme at ESI, Vienna, where part of
this work was done.

\end{document}